\documentclass{article}

\usepackage{amsthm,amssymb,amsmath}
\usepackage{graphicx}
\usepackage{epsfig}
\usepackage{cite}

\newcommand{\mH}{m_{H^0}}
\newcommand{\sigmav}{\sigma\mathrm{v}}
\newcommand{\gev}{\mathrm{GeV}}
\newcommand{\dma}{\Delta m_{A^0}}
\newcommand{\dmc}{\Delta m_{H^+}}
\newcommand{\gsim}{\lower .75ex \hbox{$\sim$} \llap{\raise .27ex \hbox{$>$}} }
\newcommand{\lsim}{\lower .75ex \hbox{$\sim$} \llap{\raise .27ex \hbox{$<$}} }

\begin{document}
\title{\bf \Large The inert doublet model of dark matter revisited}

\author{Laura Lopez Honorez\\
\small Departamento de F\'{i}sica Te\'orica C-XI and Instituto de F\'{i}sica Te\'orica UAM-CSIC, \\
\small Universidad Aut\'onoma de Madrid, Cantoblanco, E-28049 Madrid, Spain\\
\small Service de Physique Th\'eorique, Universit\'e Libre de Bruxelles, 1050 Brussels, Belgium\\
\and
Carlos E. Yaguna\\
\small Departamento de F\'{i}sica Te\'orica C-XI and Instituto de F\'{i}sica Te\'orica UAM-CSIC,\\
\small Universidad Aut\'onoma de Madrid, Cantoblanco, E-28049 Madrid, Spain
}

\date{}
\maketitle
\begin{abstract}
The inert doublet model, a minimal extension of the Standard Model
by a second higgs doublet with no direct couplings to quarks or
leptons, is one of the simplest scenarios that can
explain the dark matter.  In this paper, we study in detail the impact
of dark matter annihilation  into  the three-body final state
$WW^*$ ($\to Wf\bar f'$) on the phenomenology of the inert doublet
model.  We find that this new annihilation mode dominates, in a
relevant portion of the  parameter space, over those into two-body
final states considered in previous analysis. As a result,  the
computation of the relic density is modified and the viable regions of the model are displaced. After obtaining the
genuine viable regions for different sets of parameters,  we compute
the   direct detection cross section of inert higgs dark matter and
find it to be  up to two orders of magnitude smaller than what is
obtained for two-body final states only. Other implications of these
results, including  the modification to the decay width of the higgs
and to the indirect detection signatures of inert higgs dark matter,
are also briefly considered.  We demonstrate, therefore, that the
annihilation into the three-body final state $WW^*$ can not be
neglected, as it has a important impact on the entire phenomenology of the inert doublet model.
\end{abstract}
\section{Introduction}
Even though dark matter accounts for about $23\%$ of the energy
density of the Universe \cite{Komatsu:2008hk}, we do not yet know what exactly it consists
of. The identification of the dark matter particle is, indeed, one of
the most challenging problems in astroparticle physics today. Over the
years, many dark matter candidates have been proposed in different
scenarios for physics beyond the standard model. Among them, the inert
higgs --the lightest odd particle of the inert doublet model-- 
has earned a special place as a representative candidate of weakly
interacting scalar dark matter.

In the inert doublet model, a higgs doublet $H_2$, odd under a new
$Z_2$ symmetry, is added to the standard model particle content.  The
lightest inert (odd) particle, $H^0$, turns out to be  stable and
hence a suitable dark matter candidate.  After
being introduced in \cite{Deshpande:1977rw}, this model has been
extensively studied in a number of recent works
\cite{Ma:2006km,Barbieri:2006dq,Majumdar:2006nt,Hambye:2007vf, LopezHonorez:2006gr,Gustafsson:2007pc,Cao:2007rm,Agrawal:2008xz, Andreas:2008xy,Lundstrom:2008ai,Nezri:2009jd,Dolle:2009ft}.

Recently, it was pointed out that the annihilation of dark matter
particles can receive large contributions from three-body final
states consisting of a real and a virtual massive particle \cite{Yaguna:2010hn,Chen:1998dp,Hosotani:2009jk}. A fact that had been overlooked,
notably, in \emph{all} previous analysis of the inert doublet
model. 
The annihilation into the three-body final state $WW^*$ ($\to Wf\bar
f'$), in particular, is expected to be important in the region $M_{H^0}\lesssim
M_W$, which has already been shown to be viable and to feature a rich and interesting 
phenomenology \cite{LopezHonorez:2006gr}.
In this paper, we revisit the inert doublet model of dark
matter in view of these new contributions to the annihilation of dark matter particles. We will see that the
inclusion of the three-body final state  $WW^*$ not only 
modifies the viable parameter space of the model, but it also changes significantly the prospects for the detection of inert higgs dark matter.

In the next section we present a brief introduction to the inert doublet model, followed, in section \ref{sec:anni}, by the calculation of the annihilation cross section of inert higgs dark matter into $WW^*$ (the analytical formulas are given in the Appendix). Section \ref{sec:sigmav} is devoted to the analysis of the dependence of the three-body annihilation rate  on the parameters of the inert doublet model. In section \ref{sec:relic}, the impact of the three-body final state on  the inert higgs relic density is studied in detail. Then, we use those results, in section \ref{sec:viable}, to derive the genuine viable parameter space of the inert doublet model. The implications of this new viable parameter space for the direct detection of inert higgs dark matter are investigated in section \ref{sec:direct}. Finally, in section \ref{sec:other}, we concisely discuss  other implications of the three-body final state regarding the indirect detection of dark matter and the decays of the higgs boson.
\section{The inert doublet model}
The inert doublet model is a simple extension of the standard model
with one additional higgs doublet $H_2$ and an unbroken $Z_2$ symmetry
under which $H_2$ is odd while all other fields are even. This
discrete symmetry prevents the direct coupling of $H_2$ to fermions and,
crucial for dark matter, guarantees the stability of the lightest
inert particle. The scalar potential of this model is given by
\begin{align}
V=&\mu_1^2|H_1|^2+\mu_2^2|H_2^2|+\lambda_1|H_1|^4+\lambda_2|H_2|^4+\lambda_3|H_1|^2|H_2|^2\nonumber\\
&+\lambda_4|H_1^\dagger H_2|^2+\frac{\lambda_5}{2}\left[(H_1^\dagger H_2)^2+\mathrm{h.c.}\right]\,,
\end{align}
where $H_1$ is the standard model higgs doublet, and $\lambda_i$ and
$\mu_i$ are real parameters. Four new physical states are obtained in
this model: two charged states, $H^\pm$, and two neutral ones, $H^0$
and $A^0$. Either of them could be dark matter. In the
following we choose $H^0$ as the lightest inert particle,
$m_{H^0}^2<m_{A^0}^2,m_{H^\pm}^2$, and, consequently, as the dark
matter candidate. After electroweak symmetry breaking, the inert scalar
masses take  the following form
\begin{align}
m_{H^\pm}^2&= \mu_2^2+\frac12\lambda_3,v^2 ,\nonumber \\
m_{H^0}^2&= \mu_2^2+\frac12(\lambda_3+\lambda_4+\lambda_5)v^2 ,\nonumber \\
m_{A^0}^2&= \mu_2^2+\frac12(\lambda_3+\lambda_4-\lambda_5)v^2\,, 
\end{align}
where $v=246$ GeV is the vacuum expectation value of $H_1$. Of pertinence to our  study is the interaction term between a pair of $H^0$ and the higgs boson, which  is proportional to
$\lambda_L=(\lambda_3+\lambda_4+\lambda_5)/2$.  In addition to it,  it is convenient to take $m_{H^0}$, $\Delta
m_{A^0}=m_{A^0}-m_{H^0}$, $\Delta m_{H^\pm}=m_{H^\pm}-m_{H^0}$, and
the higgs  mass, $m_h$,  as the remaining free parameters of the inert doublet model.

In our analysis, we take into account all the known theoretical and experimental constraints on this model --see  \cite{Barbieri:2006dq} and
\cite{LopezHonorez:2006gr}. The requirement of vacuum stability imposes
\begin{equation}
\lambda_1,\lambda_2 > 0\,,\qquad 
\lambda_3, \lambda_3+\lambda_4-|\lambda_5|>-2\sqrt{\lambda_1\lambda_2}\,,
\end{equation}
whereas the precise determination of the $Z$ decay width at LEP requires that $m_{A^0}+m_{H^0}>M_Z$ and that $m_{H^+}>M_Z/2$. Because of the specific decay modes considered in the analysis, the bound  $m_{H^+}>79.3\,\gev$ on the mass of a charged higgs from LEP~\cite{Amsler:2008zzb} can not be applied to this model.  In~\cite{Pierce:2007ut}, the  constraint $m_{H^+}\gsim 70-90$ GeV was derived using  the results of the OPAL collaboration~\cite{Abbiendi:2003sc}. Some regions in the plane ($m_{H^0},m_{A^0}$) are also constrained by LEP II data, see \cite{Lundstrom:2008ai}. Additionally, the inert doublet, $H_2$,  contributes to electroweak precision observables such as $S$ and $T$. For the range of parameters we  consider, however, compatibility with present data is easily achieved.  Finally, from  section \ref{sec:viable}  on, we require also that the relic density of inert higgs dark matter be compatible with the observed dark matter density.

In previous works \cite{Barbieri:2006dq,Majumdar:2006nt,
LopezHonorez:2006gr,Gustafsson:2007pc,
Andreas:2008xy,Agrawal:2008xz,Andreas:2009hj,Nezri:2009jd,Cao:2007rm, Lundstrom:2008ai,Dolle:2009ft} several aspects of this model, including constraints from present data and prospects for dark matter detection, were studied.  It turns out that  the dark matter constraint
can only be satisfied  for restricted values of $m_{H^0}$. Three
viable regions can be distinguished:  a small mass regime with
$m_{H^0}\sim 8$ GeV~\cite{Hambye:2007vf,Andreas:2008xy}, a
large mass regime with $m_{H_0} > 500$
GeV~\cite{Cirelli:2005uq,LopezHonorez:2006gr,Hambye:2009pw} and an
intermediate mass regime ($m_{H_0} \lesssim M_W$)~\cite{Barbieri:2006dq,LopezHonorez:2006gr}.  This intermediate mass regime gives rise to a very rich phenomenology \cite{LopezHonorez:2006gr}, with significant direct detection  cross section and good indirect detection prospects
via gamma rays~\cite{Gustafsson:2007pc}.  It also coincides with the mass range where the annihilation into the three-body final state $WW^*$, a process not considered in any previous work on this model, is expected to be particularly relevant. In this  paper, we take into account, for the first time, the impact of the $WW^*$ final state on the phenomenology of the inert doublet model.
\section{$H^0H^0$ annihilation into $WW^*$ }
\label{sec:anni}

It has been recently emphasized that dark matter annihilations could
receive large additional contributions from three-body final states
consisting of a real and a virtual massive particle
\cite{Yaguna:2010hn}. The inert higgs dark matter, in particular, could annihilate into
 $WW^*$, $ZZ^*$, $hh^*$, and  $t\bar t^*$. The latter two, however,
are certainly irrelevant. The branching ratio into them is small and,
in any case, the region where they could give a significant
contribution (for $m_{H^0}$ just below the $h$ and $t$ thresholds) is
not consistent with the relic density constraint. In that region, the
annihilation into $W^+W^-$ and $ZZ$ is so effective that it drives the
thermal abundance of inert higgs dark matter well below the WMAP
measurement. Including the additional annihilation into $hh^*$ and
$t\bar t^*$ would, at most, yield an even smaller relic density. That
leaves us only with the gauge bosons three-body final states. They seem
promising because the region below the $W$ and $Z$ thresholds
partially overlaps with the intermediate viable region of the inert
doublet model, $M_{H^0}<M_W$. Between the two, we expect the $WW^*$
final state to be more important, for we know that its real
counterpart dominates the annihilation branching ratio in the region
$\mH > M_W,M_Z$. Besides, the virtual $Z$ is farther off-shell than the
virtual $W$  within the viable region of the model,
$\mH\lesssim M_W$, giving an additional suppression. It turns out that the $ZZ^*$ final state is actually negligible with respect to  $WW^*$ over the
entire viable parameter space of the inert doublet model. Thus, the only three-body final state that can modify the phenomenology of the inert double model is $WW^*$.  
\begin{figure}[tb]
\begin{center} 
\begin{tabular}{ccc}
\includegraphics[scale=0.5]{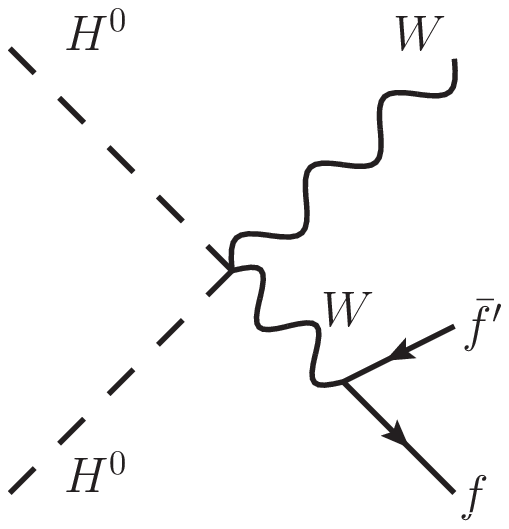} & \includegraphics[scale=0.5]{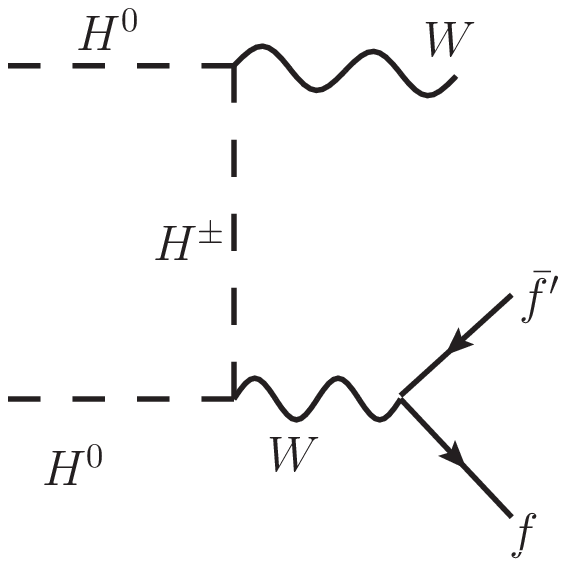} & \includegraphics[scale=0.5]{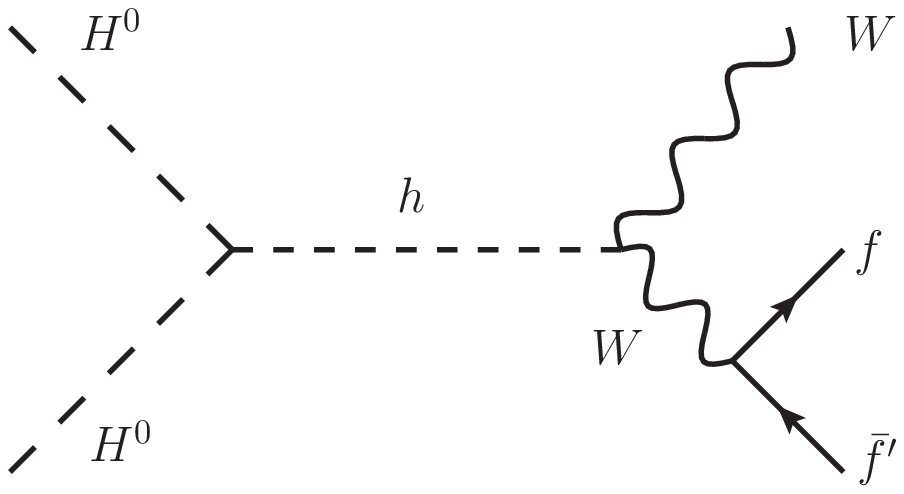}
\end{tabular}
\caption{The Feynman diagrams that contribute, in the unitary gauge, to the $H^0H^0$ annihilation into the  three-body final state $WW^*\to Wf\bar f'$ within the inert doublet model. For the $H^+$-mediated diagram the exchange diagram (not shown) must also be taken into account. \label{fig:diainert}}
\end{center}
\end{figure}

Three different diagrams contribute to the annihilation of dark matter into $WW^*$ ($H^0H^0\to WW^*\to W f\bar f'$) in the inert doublet model
--see figure \ref{fig:diainert}. The amplitude of the first diagram
depends only on gauge parameters whereas the second and the third also
depend respectively on $m_{H^+}$ and on $\lambda_L$ and $m_h$.  The contribution from the direct annihilation diagram (left diagram) is usually dominant whereas the one from the $H^\pm$-mediated diagram is typically small. The higgs-mediated contribution can be important, particularly close to the higgs resonance, where it becomes dominant. $\sigma(H^0H^0\to WW^*)$ is of utmost importance in
our discussion, as it enters explicitly into the indirect detection
rates and into the computation of the relic density. The analytical
result for this cross section can be found  in the Appendix
\ref{ap:1}.   $\sigma(H^0H^0\to WW^*)$ depends very weakly on
$m_{A^0}$ (only through the higgs width), and on $m_{H^+}$
(the $H^+$ mediated diagram is suppressed by the t-(u-)channel propagator). So, we will study its dependence on $m_{H^0}$, $\lambda_L$ (sign and magnitude), and $m_h$.  
\begin{figure}[tb]
\begin{center} 
\begin{tabular}{cc}
\includegraphics[scale=0.26]{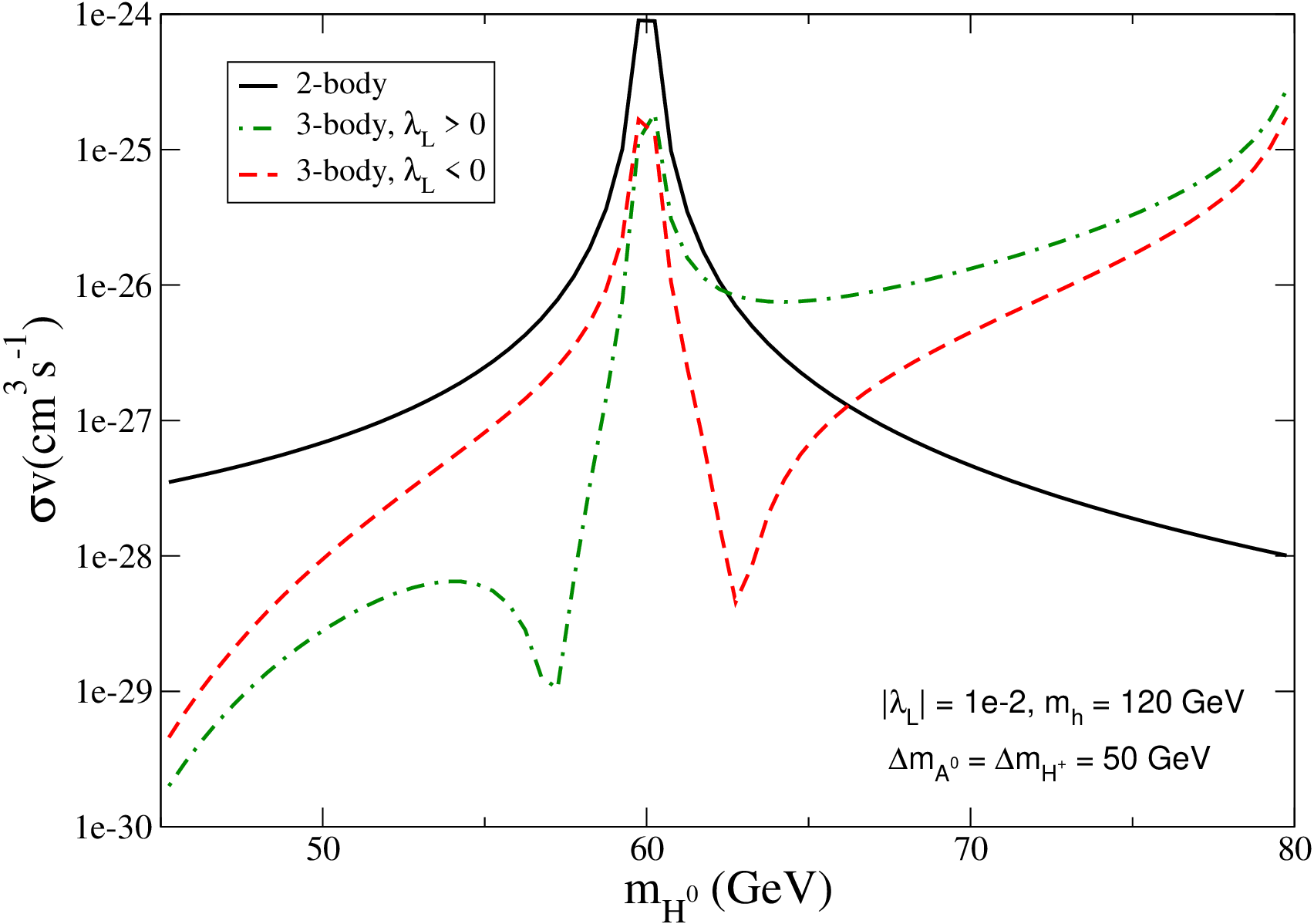} & \includegraphics[scale=0.26]{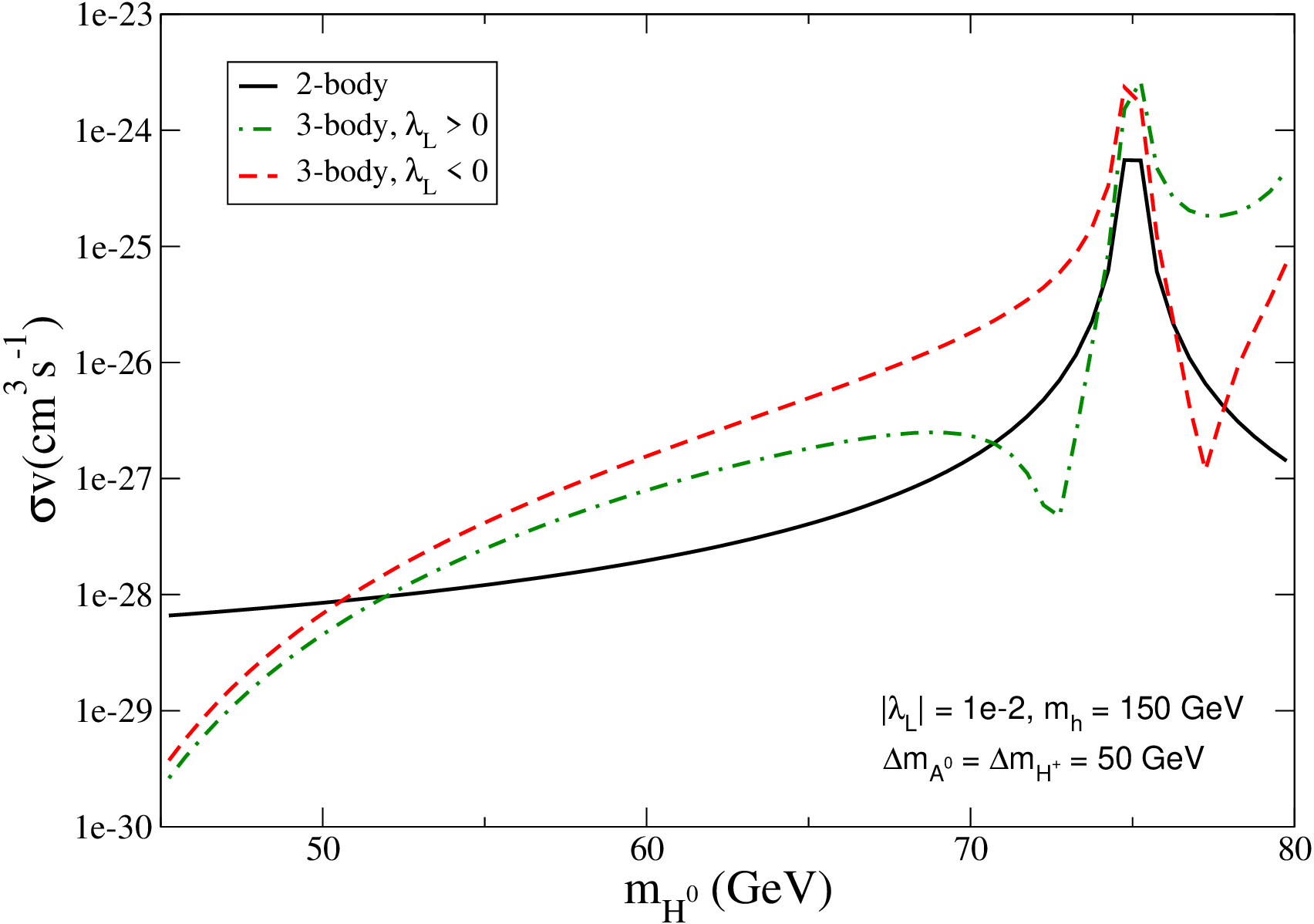}
\end{tabular}
\caption{Comparison between the three-body and the two-body
  annihilation rate, $\sigma\mathrm{v}$, as a function of the dark
  matter mass for the two possible signs of $\lambda_L$. 
In the left panel $m_h=120$ GeV whereas in the right panel $m_h=150$ GeV. The other parameters were taken as  $\Delta m_{A^0}=\Delta m_{H^\pm}=50$ GeV, $|\lambda_L|=10^{-2}$.\label{fig:svvaluelam}}
\end{center}
\end{figure}

The
two-body annihilation rate, on the other hand, is determined by
higgs-mediated processes into light fermions. So, it is proportional to $\lambda_L^2$ and it is dominated by the $b\bar b$ final state. In spite of being
formally of higher order,  the three-body process can compete with the
two-body ones thanks to the Yukawa suppression present in the latter
and to the large multiplicity  of final states associated with $WW^*$ ($\to \sum_f Wf\bar f'$). To check that it is indeed the
case, we must compute  the three-body annihilation cross section,
$\sigma(H^0H^0\to WW^*\to W f\bar f')$, and compare it with the
two-body  one. 

To begin with, let us study the behaviour of the annihilation rate at low velocities, $\sigma \text{v}$, with respect to the parameters of
the model. Figure \ref{fig:svvaluelam} compares the two-body and the
three-body ($WW^*$) annihilation rate as a function of $\mH$
for two different higgs masses, $120~\gev$ (left panel) and $150~\gev$
(right panel), and the two possible signs of $\lambda_L$. From the left panel, we see that, as expected,  the three-body cross section  generically increases as $\mH$ gets closer to $M_W$. The atypical behaviour observed around the higgs resonance, $\mH\sim 60~\gev$, is the result of the interference between the purely gauge diagram and the higgs mediated diagram, as explained in Appendix \ref{ap:2}. Because of such interference, the three-body cross section for $\lambda_L>0$ (dash-dotted line) is larger than that for $\lambda_L<0$ (dashed line) above the higgs resonance but smaller than it below the resonance. In any case, the crucial point for us is that the three-body cross section is not negligible at all. It becomes larger than  the two-body one for $\mH\gtrsim 62~\gev ~\mathrm{or}~ 67~\gev$ depending on the sign of $\lambda_L$. 

For $m_h=150~\gev$ (figure \ref{fig:svvaluelam}, right panel), the
effect is even more pronounced. In this case, the three-body cross
section dominates the annihilation rate in almost the whole range
$\mH\gtrsim 50~\gev$ --a fact  partially due to the suppression of the two-body annihilation that is expected for a higher higgs mass.  We see that the only regions where the two-body cross
section is larger are two narrow mass intervals around the resonance
where the interference effects between the higgs and pure gauge
contribution suppress the three-body cross section. Also notice that
right at the resonance, the two-body annihilation rate is larger than
the three-body one for $m_h=120~\gev$ whereas it is the other way
around for $m_h=150~\gev$. This is in agreement with the known result
that in the Standard Model a $120~\gev$ higgs boson decays dominantly
into two-body final states whereas a $150~\gev$ higgs boson decays
mainly into three-body final states. For even higher values of the higgs mass, the differences between the $\lambda_L>0$ and
$\lambda_L<0$ cases  will tend to fade out as the higgs resonance
moves further away from the relevant parameter space.  We have thus illustrated, via figure  \ref{fig:svvaluelam}, the importance  of $\sigma(H^0H^0\to WW^*)\mathrm{v}$ and its dependence on $m_{H^0}$, $m_h$ and the sign of $\lambda_L$.

\section{Effects on $\sigma v$ - a systematic analysis}
\label{sec:sigmav}

\begin{figure}[tb]
\begin{center} 
\includegraphics[scale=0.35]{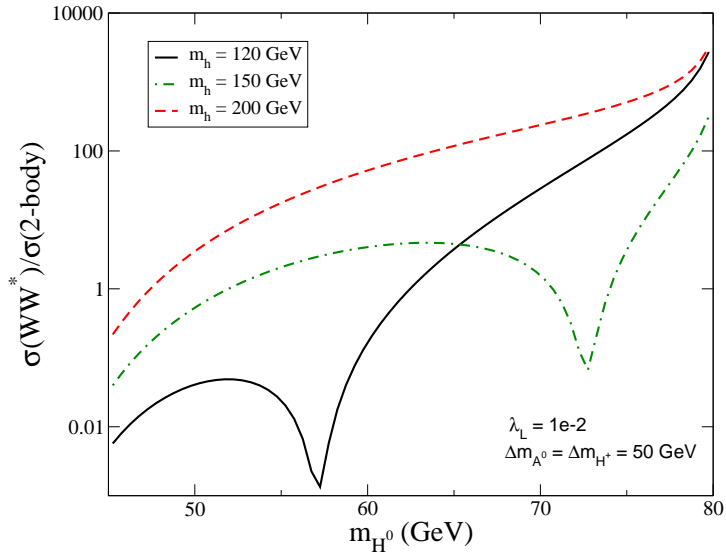}
\caption{This figure illustrates the dependence on $m_h$ of the ratio
  between the three-body and the two-body  annihilation rate. For the other parameters, we take  $\Delta m_{A^0}=\Delta m_{H^\pm}=50$ GeV and $\lambda_L=10^{-2}$.\label{fig:sigmav}}
\end{center}
\end{figure}
With the goal of understanding the relevance of the three-body final
state, in this section we perform a systematic analysis of the ratio
between the three-body cross section $\sigma(H^0H^0\to WW^*)$
(sometimes denoted simply as $\sigma(WW^*)$) and  the two-body one
$\sigma(H^0H^0\to \sum_f f\bar f)$ (denoted also as
$\sigma(\text{2-body})$) within the inert doublet model. For the
higgs mass, we consider three typical values compatible with electroweak precision data: $120$, $150$, and $200\,\gev$.

\begin{figure}[tb]
\begin{center} 
\includegraphics[scale=0.35]{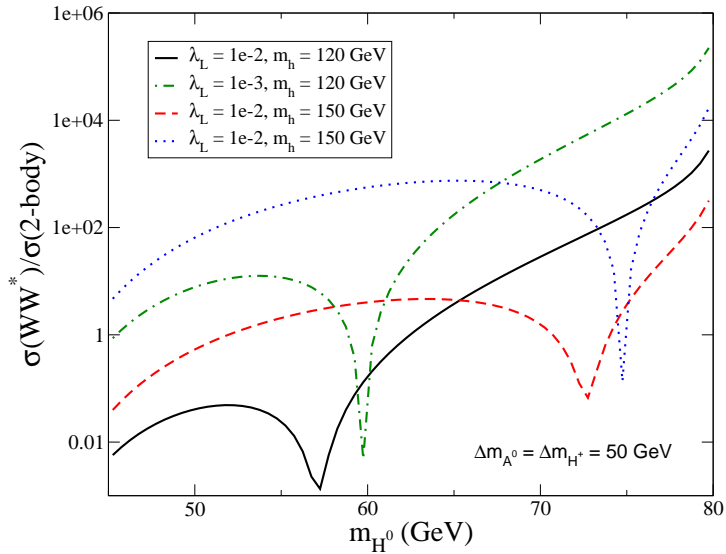}
\caption{This figure illustrates the dependence on $\lambda_L$ of the ratio
  between the three-body and the two-body  annihilation rate. It shows $\sigma(\text{3-body})/\sigma(\text{2-body})$ as a function of $m_{H^0}$ for two values of $\lambda_L$ ($10^{-2}$ and $10^{-3}$) and two different higgs masses ($120$ and $150\,\gev$). For the other parameters, we take  $\Delta
  m_{A^0}=\Delta m_{H^\pm}=50$ GeV.\label{fig:svlambda}}
\end{center}
\end{figure}

Figure \ref{fig:sigmav} displays the ratio of cross sections as a
function of $\mH$  for different values of the higgs mass. For this
figure $\lambda_L$ was set to $10^{-2}$ whereas $\dma$ and $\dmc$,
which hardly affect the results, were set to $50~\gev$. From the
figure we see that, the ratio  is larger for $m_h=200~\gev$ over the
whole range of $\mH$. For that higgs mass, the ratio increases with
$\mH$ and it is  larger than $1$ in the entire range
$M_W>\mH>50~\gev$,  implying a dominance of the three-body final state
over the two-body ones. For the other two higgs masses, we see that
the ratio tends to increase with $\mH$ but it features a narrow dip
before reaching the higgs resonance. For $m_h=120~\gev$ the three-body
final state dominates the cross section for $\mH\gtrsim 62~\gev$
whereas for $m_h=150~\gev$ it dominates it over the range $\mH\gtrsim
52~\gev$ except for a small mass range around $73~\gev$. Independently of the higgs mass, the ratio can be larger than $100$ close to the $W$ threshold, indicating a strong dominance of the three-body final states in that region. 

\begin{figure}[tb]
\begin{center} 
\includegraphics[scale=0.35]{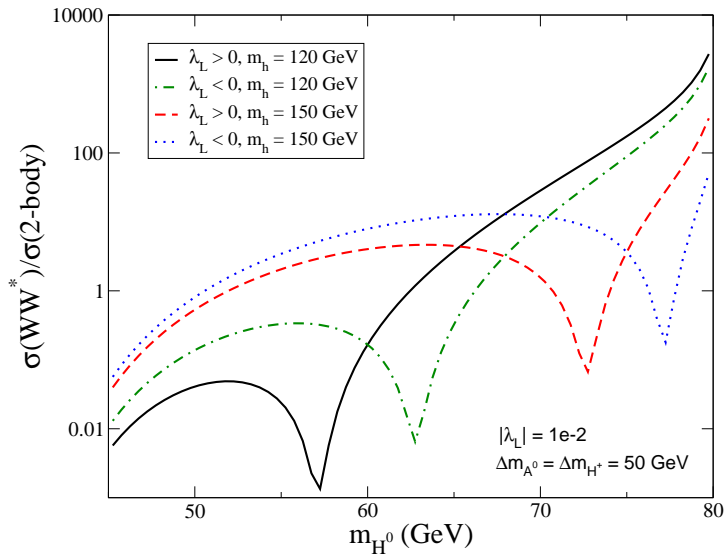}
\caption{
This figure illustrates the dependence on the sign of $\lambda_L$ of the ratio
  between the three-body and the two-body  annihilation rate. It shows $\sigma(\text{3-body})/\sigma(\text{2-body})$ as a function of $m_{H^0}$ for the possible signs of $\lambda_L$ and two different higgs masses ($120$ and $150\,\gev$). For
  the other parameters, we take  $\Delta m_{A^0}=\Delta m_{H^\pm}=50$
  GeV and $|\lambda_L|=10^{-2}$.\label{fig:svsignlam}
}
\end{center}
\end{figure}

The annihilation cross section into $WW^*$ is affected by  $\lambda_L$ via the higgs mediated diagram. This dependence is illustrated in figure \ref{fig:svlambda}, which shows as a function of $\mH$, the ratio of cross sections for two different values of $\lambda_L$ ($10^{-2}$ and $10^{-3}$) and two different higgs masses ($120~\gev$ and $150~\gev$). As expected, a smaller $\lambda_L$ suppresses the two-body cross section, which being  higgs mediated is proportional to $\lambda_L^2$, increasing the relevance of the three-body final state. In fact, the ratio is about two orders of magnitude larger for $\lambda_L=10^{-3}$ than for $\lambda_L=10^{-2}$. As a result, for $\lambda_L=10^{-3}$ the three-body cross section is larger than the two body one essentially over the entire $\mH$ range we consider, and the ratio can reach values above $10^4$ close to the $W$ threshold. Another feature observed in the figure is the displacement of the position of the dip, which due to the smaller interference effects moves closer to the resonance for smaller values of $\lambda_L$ --see appendix \ref{ap:2}.

As already shown in figure~\ref{fig:svvaluelam}, the three-body cross
section depends also on the sign of $\lambda_L$, due to the
interference between the higgs mediated amplitude, which goes like
$\lambda_L$, and the other two contributions. Figure
\ref{fig:svsignlam} shows the ratio of cross sections as a function of
$\mH$ for the two possible signs of $\lambda_L$ and two different
higgs masses: $120$, $150\,\gev$. The other parameters were taken as
$|\lambda_L|=10^{-2}$, $\dma=\dmc=50~\gev$. The sign of $\lambda_L$
clearly influences the three-body cross section, particularly around
the higgs resonance, where the interference is stronger. Notice that
the dip moves across the resonance when $\lambda_L$ changes
sign. Moreover, the curves for  $\lambda_L>0$ and  $\lambda_L<0$ cross
each other exactly at the resonance, as expected. Indeed at that point, the
annihilation cross section is entirely determined by the higgs
mediated diagram, so $\sigma(WW^*)\propto \lambda_L^2$ there. 
Notice that the ratio is larger for $\lambda_L>0$  than for $\lambda_L<0$ above the resonance, whereas it is the other way around below the resonance. Thus, the largest effect will be obtained for $\lambda_L>0$ if $m_h=120\,\gev$ but for $\lambda_L<0$ if $m_h=150\,\gev$. 

This detailed study of $\sigma(H^0H^0\to WW^*)/\sigma(H^0H^0\to f\bar
f')$ clearly demonstrates that the relevance of the annihilation into
the three-body final state $WW^*$ is a generic feature of the inert
doublet model. No fine-tuning is necessary to find  large effects. In
the next section we investigate the implications  of this new process on the calculation of the inert higgs relic density.
 
\section{Effects on the relic density}
\label{sec:relic}
To compute the dark matter relic density, one must solve the following Boltzmann equation,
\begin{equation}
\frac{dY}{dT}=\sqrt{\frac{\pi g_*(T)}{45}}M_p\langle\sigmav\rangle (Y(T)^2-Y_{eq}(T)^2)\,,
\label{boltzmann}
\end{equation}
where  $Y(T)$ is the dark matter abundance,  defined as the number
density divided by the entropy density, $g_*$ is an effective number
of degrees of freedom, $M_p$ is the Planck mass, and $Y_{eq}(T)$ is
the equilibrium thermal abundance. $\langle\sigmav\rangle$ is the
thermally averaged annihilation cross section, which must include all
relevant annihilation and coannihilation processes. It corresponds to
\begin{equation}
\langle\sigmav\rangle=\frac{\sum_{i,j}g_ig_j \int_{(m_i+m_j)^2} ds\sqrt{s}K_1(\sqrt s/T)p_{ij}^2\sum_{k,l}\sigma_{ij;kl}(s)}{2T\left(\sum_ig_im_i^2K_2(m_i/T)\right)^2}\,,
\label{thermalav}
\end{equation}
where $i,j$ run over all annihilating and coannihilating particles,
$g_i$ is the number of degrees of freedom of particle $i$, $m_i$ is
its mass, $p_{ij}$ is the momentum of the incoming particles in the
center of mass frame, and  $\sigma_{ij;kl}$ is the total annihilation
cross section of particles $i,j$ into Standard Model particles
$k,l$. Integrating equation (\ref{boltzmann}) down to today's temperature $T=T_0$ leads to the present dark matter abundance, $Y(T_0)$. From it, the  relic density can be obtained as
\begin{equation}
\Omega h^2= 2.74\times 10^8 \frac{M_{dm}}{\gev}Y(T_0)\,,
\label{omega}
\end{equation}
where $M_{dm}$ is the dark matter particle mass. This  standard approach  for the calculation of the relic density  has been implemented in publicly available software such as DarkSUSY\cite{gondolo-2004-0407,DSweb} and micrOMEGAs\cite{Belanger:2008sj,Belanger:2006is,Belanger:2004yn,Belanger:2001fz}. 

To properly compute the inert higgs relic density we need to include in equation
(\ref{thermalav}), in addition to the usual
two-body annihilation and coannihilation processes, the annihilation
into the three-body final state $WW^*$. That is the only modification
we need to make to the above procedure --equations (\ref{boltzmann})
and (\ref{omega}) remain the same. For our calculations, we have used
a modified version of micrOMEGAs, in which we incorporated the
annihilation into  the three-body final state $WW^*$ in the evaluation
of~(\ref{thermalav}). That way, we can  accurately compute the relic density of inert higgs dark matter including 3-body final states.

In the following we compare the relic density obtained for two-body final states only (denoted as $\Omega(\text{2-body})$) with that predicted including also the final state $WW^*$ (denoted as $\Omega(\text{3-body})$ and referred to as the 3-body relic density) for different values of the parameters of the inert doublet model. 

\begin{figure}[tb]
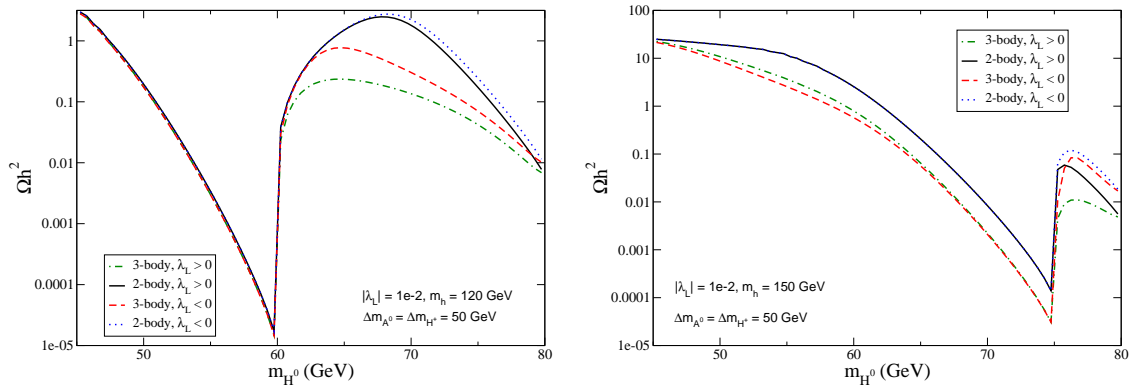

\begin{center} 
\begin{tabular}{cc}
\includegraphics[scale=0.25]{omvaluelam.eps} & \includegraphics[scale=0.25]{omvaluelam2.eps}
\end{tabular}
\caption{Comparison between the 3-body and the 2-body
  relic density, $\Omega h^2$, as a function of the dark matter mass for the two possible signs of $\lambda_L$. In the left panel $m_h=120$ GeV whereas in the right panel $m_h=150$ GeV. The other parameters were taken as  $\Delta m_{A^0}=\Delta m_{H^\pm}=50$ GeV, $|\lambda_L|=10^{-2}$.\label{fig:omvaluelam}}
\end{center}
\end{figure}

Figure \ref{fig:omvaluelam} shows the relic density derived including
only  2-body processes (denoted 2-body)  and including 2-body and
3-body processes relic density (denoted 3-body) as a function of $\mH$
for the two possible signs of $\lambda_L$ and two different higgs
masses: $120~\gev$ (left panel) and $150~\gev$ (right panel). The
remaining parameters were taken as $|\lambda_L|=10^{-2}$,
$\dma=\dmc=50~\gev$. First of all, notice that the two-body relic
density (solid and dotted lines) depends weakly on the sign of
$\lambda_L$, and only for $\mH$ close to $M_W$. This behavior is to be
expected as it is the annihilation into $W^+W^-$ (both  being real
particles) that brings such a dependence into play, and it is only for
$\mH$ close to $M_W$ that such annihilation can take place in the
early Universe. The 3-body relic density is observed to be always
equal or smaller than the two-body one --as anticipated-- and to
depend  on the sign of $\lambda_L$. For $m_h=120~\gev$ (left panel)
the effect of the 3-body relic density is  significant above the
resonance, and the predicted relic density is smaller for
$\lambda_L>0$ (dash-dotted line) than for $\lambda_L<0$ (dashed
line). For $\mH$ close to $M_W$, the dark matter particles may have
enough kinetic energy in the early Universe to  annihilate into
$W^+W^-$; consequently, the effect of the three-body final state
becomes less relevant in that region. For $m_h=150~\gev$ (right
panel), the 3-body relic density is significantly smaller than the
two-body one over the entire mass range we consider. Regarding the
sign of $\lambda_L$, both signs give approximately the same relic
density below the resonance and differ from each other above it. These
two figures demonstrate that the three-body final state $WW^*$ affects
in a relevant way the predicted relic density of inert higgs dark
matter.

\begin{figure}[t]
\begin{center} 
\includegraphics[scale=0.35]{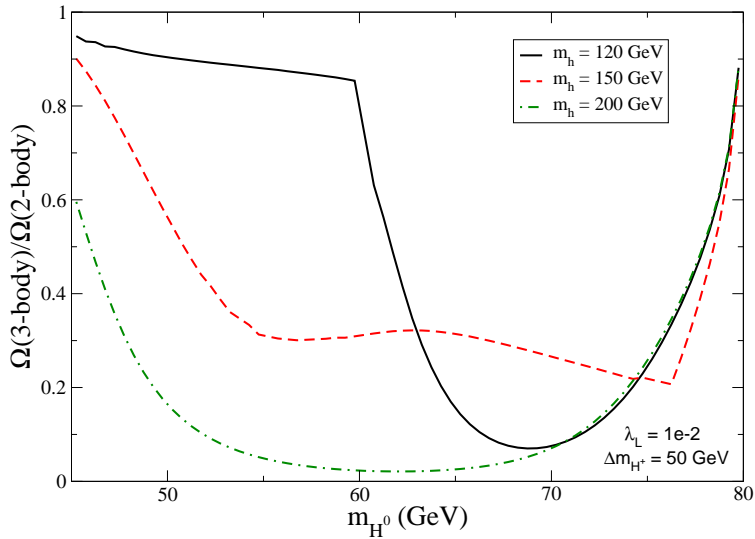}
\caption{Figure illustrating the dependence in the higgs mass of the ratio between the
  relic density including the three-body final state and the relic
  density for two-body final states as a function of the dark matter
  mass. The other parameters were taken as  $\Delta m_{A^0}=\Delta m_{H^\pm}=50$ GeV and $\lambda_L=10^{-2}$.\label{fig:ommh}}
\end{center}
\end{figure}

To better illustrate the effect of the three-body final state, we next study  the ratio between the three-body and the two-body relic density for different sets of parameters. Figure \ref{fig:ommh} shows this ratio as a function of $\mH$ for different values of the higgs mass. A ratio equal to $1$ means that the three-body process gives a negligible correction to the calculation of the relic density. Clearly, that is not the case. The ratio tends to $1$ for $\mH$ close to $M_W$, where the annihilation into $W^+W^-$ is efficient, and for $\mH\ll M_W$, where the three-body annihilation is suppressed, but in the intermediate region  the three-body final state plays a major role, giving rise to a correct relic density significantly smaller than the two-body one. An effect that, as observed in the figure, is present independently of the higgs mass --although its precise magnitude will certainly depend on $m_h$. Notice, from  figure \ref{fig:ommh}, that the  two-body
approximation may overestimate the predicted relic density by more than one order of
magnitude. Moreover, a significant deviation from the two-body result can take place over a wide range of $\mH$;  depending on the higgs mass, it could extend more than $30$ GeV below $M_W$. 
\begin{figure}[t]
\begin{center} 
\includegraphics[scale=0.35]{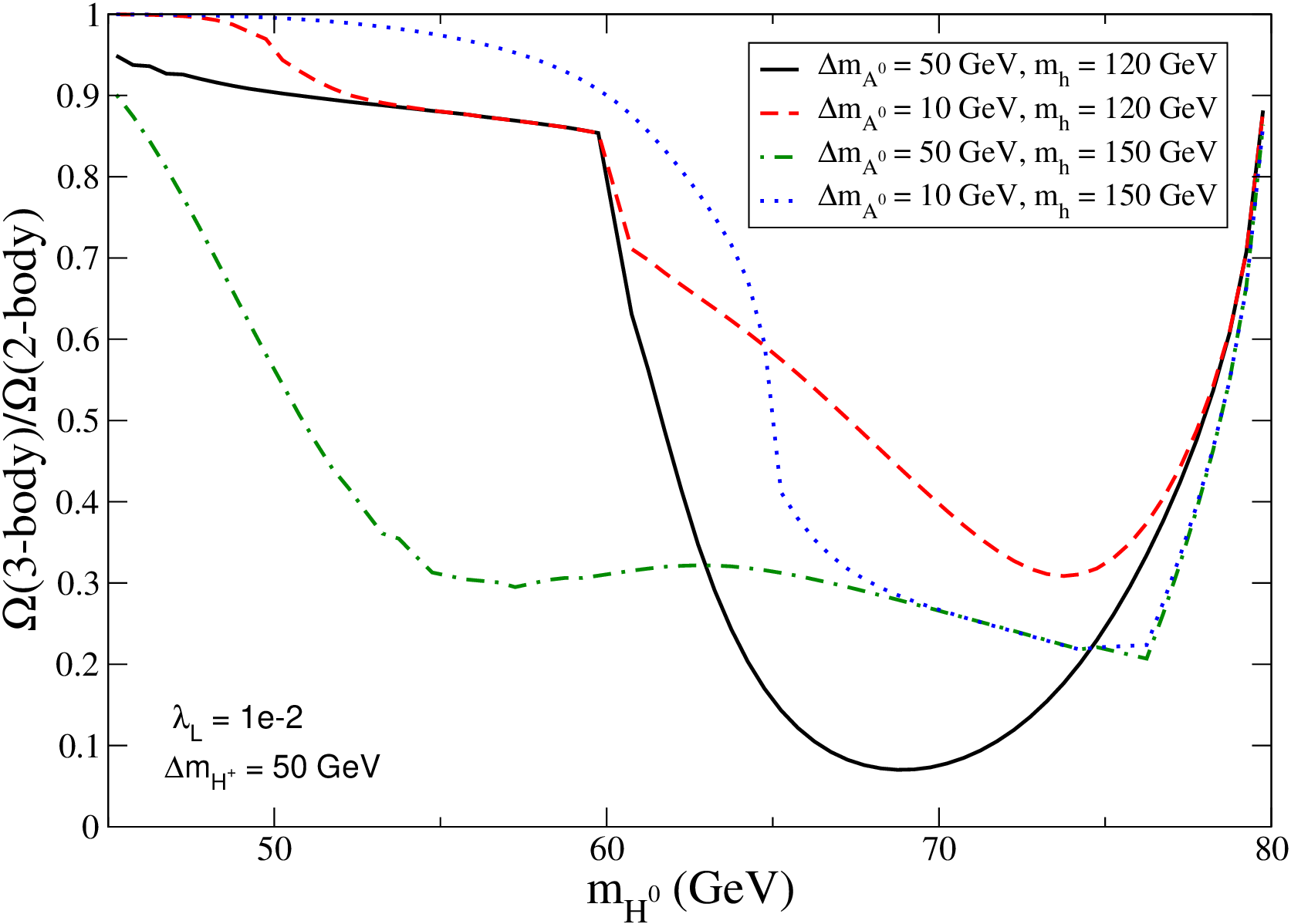}
\caption{Figure illustrating the dependence on $\dma$ of the
  ratio between the relic density including the three-body final state
  and the relic density for two-body final states only. $\dma$ affects $\Omega h^2$ only through coannihilations effects, which are important for small mass splittings ($\dma=10\,\gev$) but not for large ones ($\dma=50\,\gev$). Two different values of the higgs mass are shown. The other parameters were taken as  $\Delta m_{H^\pm}=50$ GeV and $\lambda_L=10^{-2}$.\label{fig:omma0}}
\end{center}
\end{figure}

The mass of the CP-odd scalar, $m_{A^0}$, may affect the $H^0$ relic
density via coannihilation processes. If the mass splitting between
$H^0$ and $A^0$, $\dma$, is small, the process $H^0A^0\to Z^*\to f\bar
f'$ increases the annihilation rate and help reduce the relic
density. Let us point out that, in the inert doublet model,  there is
no need to consider possible coannihilations into three-body final
states because the two-body ones, being gauge processes, are unsuppressed. Since coannihilations increase the annihilation rate, they
are expected to reduce the importance of the three-body final
state. That is exactly what is seen in Figure \ref{fig:omma0}, which
shows the effect of a smaller $\dma$ on the  relic density. It
displays the ratio between the 3-body and the 2-body relic density
for two  values of $\dma$ and two different higgs masses. When
$\dma=10~\gev$,  coannihilations are important (dotted- and
dashed-lines), and the three-body final state is less
relevant --the ratio is closer to $1$-- than for $\dma=50~\gev$  when
coannihilations are suppressed (solid- and dashed-dotted lines). For $m_h=150~\gev$
the deviation due to the different values of $\dma$ is significant for $\mH$ below $65~\gev$ whereas for $m_h=120~\gev$ it is so for $\mH$ between $60~\gev$ and $75~\gev$. Notice also that close to the higgs resonance, the $H^0H^0$ annihilation rate is enhanced, making coannihilation effects negligible. That is why the curves for $\dma=10~\gev$ and $\dma=50~\gev$ coincide in the region $\mH\lesssim m_h/2$. Although coannihilation effects slightly reduce its relevance, the effect of the three-body final state remains significant over the parameter space of the inert doublet model.

\section{The genuine viable parameter space}
\label{sec:viable}
\begin{figure}[t]
\begin{center} 
\includegraphics[scale=0.35]{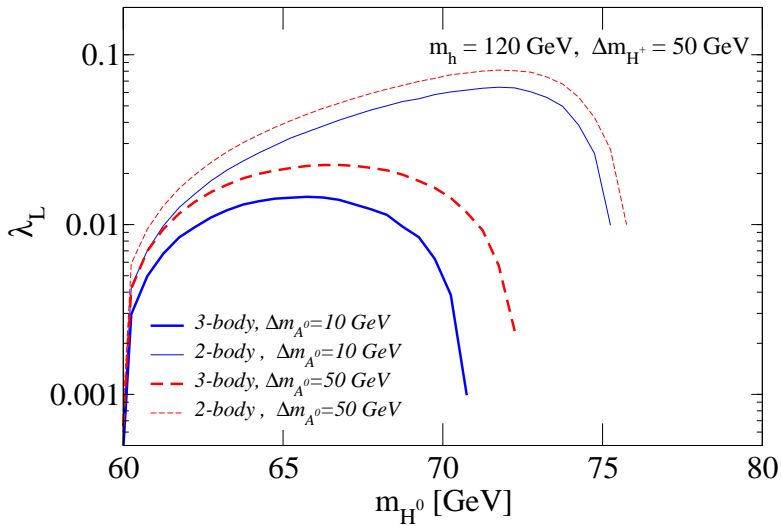}
\caption{The viable parameter space for $m_h=120\mathrm{GeV}$. Along the lines $\Omega h^2=0.11$. The thick lines are the result including the final state $WW^*$, the thin lines correspond to 2-body final states only. \label{fig:lammh120}}
\end{center}
\end{figure}

Instead of computing the dark matter density for a given set of
parameters, we are usually interested in using the relic density as a
constraint on the parameter space of the model. The viable parameter
space is determined by  requiring that  the predicted relic abundance be
compatible with the observed density of dark matter
\cite{Komatsu:2008hk}. For the inert doublet model, this viable
parameter space had been obtained in previous works, but using the
two-body relic density, which, as shown in the previous section,  is
not a good approximation in the intermediate mass regime. In this section,  the genuine viable parameter space of the inert doublet model,  obtained by including the three-body annihilation into the computation of the relic density, is derived. We study in detail its dependence on the parameters of the model and we demonstrate that it is significantly different from that found for two-body annihilations.

For definiteness, we  focus on the following  three interesting cases: $m_h=120\,\gev$ with $\lambda_L>0$,   $m_h=150\,\gev$ with $\lambda_L<0$, and $m_h=200\,\gev$ with $\lambda_L<0$.  Figure \ref{fig:lammh120} shows the viable
parameter space of the intermediate mass range of the inert dark matter model in the plane
($\lambda_L,m_{H^0}$) for $m_h=120$ GeV, $\Delta m_{H^\pm}=50$ GeV, and two
different values of $\Delta m_{A_0}$, $10$ GeV and $50$ GeV. These two values of $\Delta m_{A_0}$ are chosen so as to indicate the possible effect of having ($\Delta m_{A_0}=10\,\gev$) or not ($\Delta m_{A_0}=50\,\gev$) significant coannihilation processes. The thin lines in this figure
correspond to the viable regions if only two-body final states are considered. The thick lines,
on the contrary, correspond to the \emph{genuine} viable regions, those obtained by taking into account two- and three-body
final states in the calculation of the relic density. Two important results are clearly observed in
 this figure. First, the value of $\lambda_L$ for a given mass may be substantially smaller
once three-body final states are taken into account. The difference between the two $\lambda_L$ associated to a fixed value of $\Delta m_{A^0}$ and $m_{H^0}$  could  amount
to one order of magnitude. Second, the viable 
parameter space shrinks toward lower masses. When only two-body final
states are considered, the maximum values of  $m_{H^0}$ allowed  are about $75$
and $76$ GeV respectively for $\Delta m_{A^0}=10$ and $50$ GeV. Once
the three-body final state is included, the maximum $m_{H^0}$ moves
respectively to about $71$ GeV and $73$ GeV. This reduction of the
viable mass range is entirely due to the effectiveness of the
three-body annihilation, which drives the relic density below the WMAP
bound for $\mH>71,73\,\gev$. In that region, no value
of $\lambda_L$ is allowed, for it is the direct annihilation (figure \ref{fig:diainert}, left diagram), a gauge process independent of $\lambda_L$, that determines the relic density. As observed in this figure,
the difference between the two-body and the three-body viable
parameter space is quite significant, independently of possible
coannihilation effects.
\begin{figure}[t]
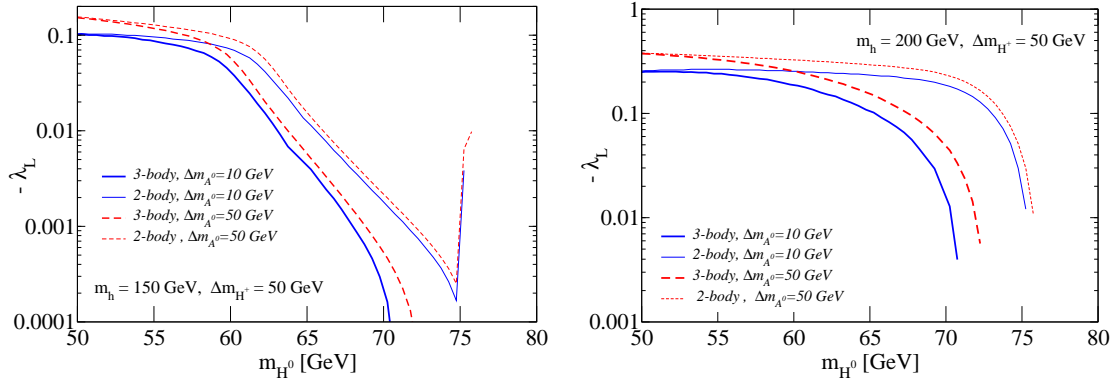

\begin{center} 
\begin{tabular}{cc}
\includegraphics[scale=0.25]{lammh150.eps} & \includegraphics[scale=0.25]{lammh200.eps}
\end{tabular}
\caption{The viable parameter space for $m_h=150\mathrm{GeV}$ (left panel) and $m_h=200\mathrm{GeV}$ (right panel). Along the lines $\Omega h^2=0.11$. The thick lines are the result including the final state $WW^*$, the thin lines correspond to 2-body final states only.\label{fig:lammh150}}
\end{center}
\end{figure}

Similar results are found for other higgs masses, as illustrated in
figure \ref{fig:lammh150} for $m_h=150~\gev$ (left panel) and
$m_h=200~\gev$ (right panel).  In both cases  $\lambda_L<0$  was considered and the conventions used were the same  as in figure
\ref{fig:lammh120}. As a consequence of the three-body final state
contribution to the annihilation rate of inert higgs dark matter, the
required value of $\lambda_L$ is smaller at any given mass, and the
maximum allowed value of $\mH$ gets reduced by several $\gev$s.

Figures \ref{fig:lammh120} and \ref{fig:lammh150} confirm that the
modification of the viable parameter space, induced by the
annihilation into the three-body final state $WW^*$, is a generic
feature of the inert doublet model. A feature that is present over a
wide range of $\mH$ independently of the other parameters of the
model. It is precisely because of this generality that the intermediate mass regime of the inert doublet model  must be
revisited, as we do in this paper, in view of these new processes.

These modifications to the viable parameters space are of crucial relevance because they affect other phenomenological aspects  of the model. In fact, the first step in the analysis of a given dark matter model is usually the determination of its viable parameter space. Once it is obtained, one can study the specific signatures or predictions of the model within such viable regions. If these are changed so are its predictions and signatures. In the next section we show that, as a result of the new viable parameter space, the direct detection cross section of inert higgs dark matter is considerably reduced.

\section{Direct detection of inert higgs dark matter}
\label{sec:direct}
\begin{figure}[t]
\begin{center} 
\includegraphics[scale=0.35]{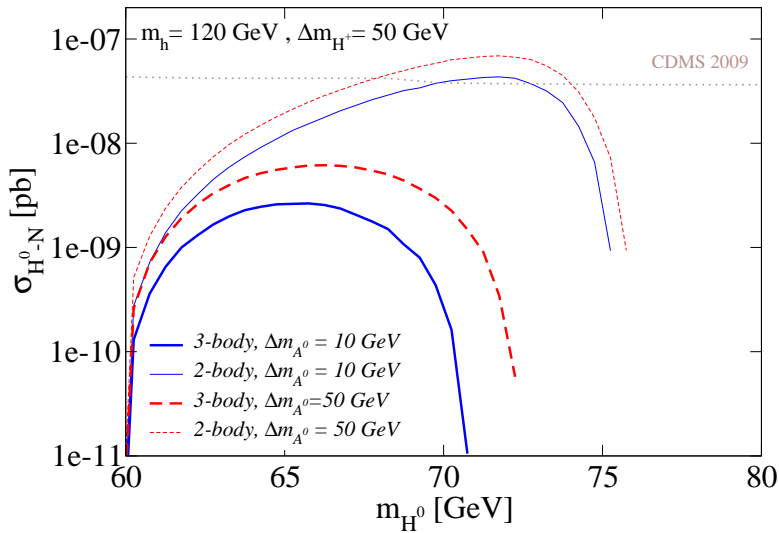}
\caption{The direct detection cross section along the viable regions
  of the inert doublet model for $m_h=120$ GeV and $\dmc=50$ GeV and
  two different values of $\dma$.\label{fig:sdd120}}
\end{center}
\end{figure}
Dark matter can be detected via elastic scattering with terrestrial detectors, the so-called direct detection method. From a particle physics point of view, the quantity that determines the direct detection rate of a dark matter particle is the dark matter-nucleon scattering cross section. In the inert higgs model, the $H^0N$  scattering
process relevant for  direct detection is higgs-mediated, with a
cross  section, $\sigma_{H^0N}$, given by 
\begin{equation}
\sigma_{H^0N}=\frac{m_r^2}{4\pi}\left( \frac{\lambda_L}{\mH m_h^2}\right)^2f^2m_N^2\,
\end{equation}
where $m_r$ is the reduced mass of the system, $m_N$ is the nucleon
mass that we took equal to the proton mass,  and $f$ is the nucleon form factor,  taken equal to  $0.3$ for the
subsequent analysis (see e.g. the discussion in \cite{Andreas:2008xy}).
Hence, $\sigma_{H^0N}$  is proportional to $\lambda_L^2$. Given the new allowed values
of $\lambda_L$ that were derived in the previous section, we foresee that  $\sigma_{H^0N}$ will be  significantly reduced with respect to the two-body result used, until now, in the literature.  
\begin{figure}[t]
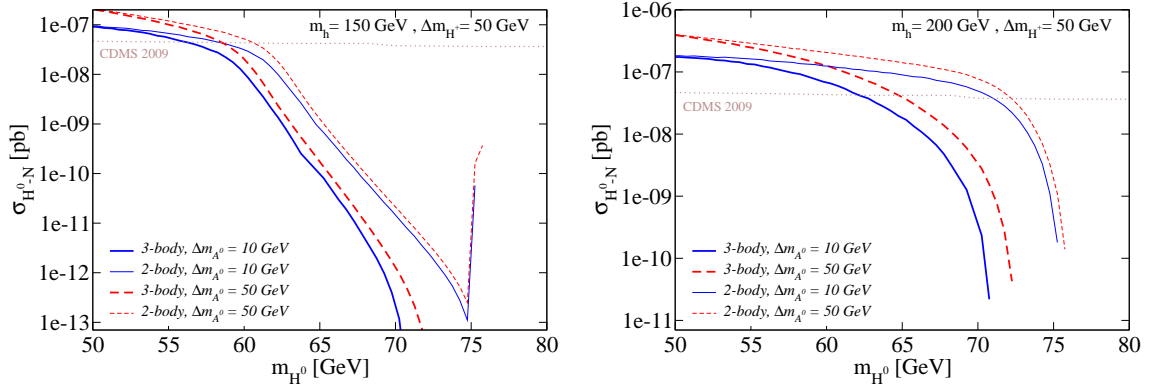

\begin{center} 
\begin{tabular}{cc}
 \includegraphics[scale=0.25]{sdd150.eps} & \includegraphics[scale=0.25]{sdd200.eps}
\end{tabular}
\caption{The direct detection cross section along the viable regions
  of the inert doublet model for $m_h=150~\gev$ (left panel) and
  $m_h=200~\gev$ (right panel). For the other parameters, we took $\dmc=50$ GeV and
  $\dma=10, 50$ GeV.\label{fig:sdd150}}
\end{center}
\end{figure}

Figure \ref{fig:sdd120} shows the prediction for $\sigma_{H^0N}$ along the viable lines of the inert doublet model for $m_h=120~\gev$. As before, thin lines correspond to the two-body result while thick lines to the three-body one. For comparison, in this figure we also show, as a dotted line, the current limit
from CDMS \cite{Ahmed:2009zw}.  Notice from the figure that the correct direct detection cross section
can be more than  two orders of magnitude smaller than the one obtained for
two-body final states. In particular, for $m_h=120\,\gev$ the genuine direct detection cross section turns out to be  well below  present bounds.

Analogous results follow also for other values of the higgs
mass. Figure \ref{fig:sdd150} shows the same cross section but for
$m_h=150\,\gev$ (left panel) and $m_h=200\,\gev$ (right panel). In
both we observe that, over a wide range of
$\mH$, the correct cross section  (denoted by 3-body)  is
significantly smaller than the one derived after including only 2-body
processes in the computation of the relic abundance (denoted by 2-body). Notice, for instance, that for $m_h=200\,\gev$ the region above
the CDMS bound moves from $\mH<71\,\gev$ to $\mH<62\,\gev$ when the
three-body final state is taken into account.

The inclusion of three-body final states is, therefore, mandatory if one wants to make meaningful
predictions on the prospects for the direct detection of inert higgs dark matter.

\section{Other implications}
\label{sec:other}
In this section we briefly address other possible implications of the
three-body process on the phenomenology of the inert doublet
model. Specifically, we show, in section~\ref{sec:ind}, that the annihilation branching ratios can be dominated by the three-body final state, with important implications for the indirect detection of dark matter, and we discuss, in section~\ref{sec:higgs}, the modifications to the decay width of the higgs boson. 
\subsection{Indirect detection}
\label{sec:ind}
\begin{figure}[t]
\begin{center} 
\includegraphics[scale=0.35]{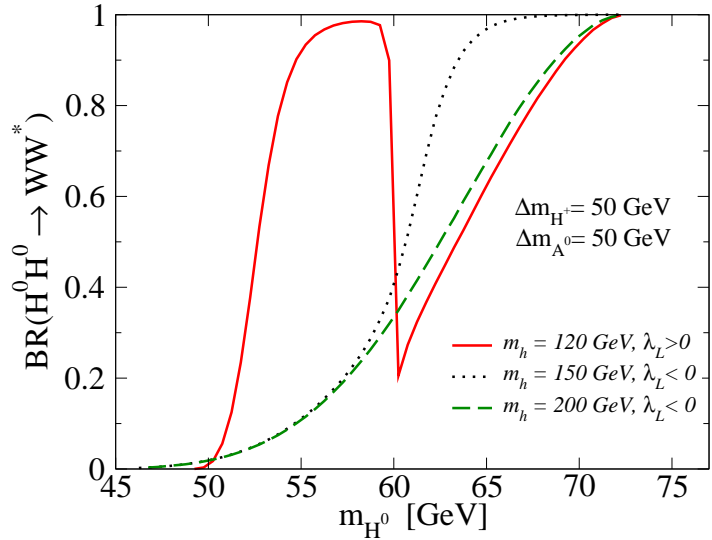}
\caption{The annihilation branching ratio into the three-body final state $WW^*$ along the viable regions of the inert doublet model.\label{fig:br}}
\end{center}
\end{figure}
The indirect detection signals of inert higgs dark matter are also
altered by  the existence of the three-body final state $WW^*$. On the
one hand,  these signals should be now computed along new regions, due
to the modified viable parameter space. On the other hand, in these
new regions the annihilation cross section and branching ratios
typically receive large corrections from the three-body final state
$WW^*$. Indeed, we already saw, in section~\ref{sec:sigmav},  that the
annihilation of inert higgs dark matter may be dominated by three-body
final states, rather than by the two-body final states considered in
previous works. As a result,  the spectrum of photons, neutrinos,
positrons and antiprotons  expected from inert higgs annihilation will
be different,  changing its indirect detection prospects. Even the
gamma ray lines from the one-loop annihilation into two photons
\cite{Gustafsson:2007pc} will be affected by these new
contributions. A detailed study of the implications of three-body
final states for the indirect detection of inert higgs dark matter is
beyond the scope of the present paper. Here, we just want to
demonstrate that, contrary to what has been assumed in earlier
analysis on indirect detection of inert higgs dark matter
\cite{Agrawal:2008xz}, $b\bar b$ is not necessarily the dominant
annihilation channel for $m_{H^0}\lsim M_W$. The three-body final state
$WW^*$ turns out to be dominant over a sizeable region of the viable
parameter space.

Figure \ref{fig:br} shows the annihilation branching ratio into the
three-body final state $WW^*$ as  a function of $\mH$ along the viable
regions of the inert doublet model. Each line corresponds to a given
value of $m_h$ and a given sign of $\lambda_L$. For this figure we set
$\dma=\dmc=50~\gev$ but the results are similar for other allowed
values. Notice that the branching into $WW^*$ is indeed significant:
it amounts to more than $10\%$ for $\mH>55~\gev$ independently of the
higgs mass. Moreover,  it  reaches values close to $1$ for
$55~\gev<\mH<60~\gev$ if $m_h=120~\gev$ and also for
$65~\gev<\mH<72~\gev$ independently of the higgs mass. In view of
these results, the indirect detection signatures of inert higgs dark
matter will have to be revised. In a future work, we plan to carry out such an analysis.
\subsection{Higgs decays}
\label{sec:higgs}
\begin{figure}[t]
\begin{center} 
\includegraphics[scale=0.35]{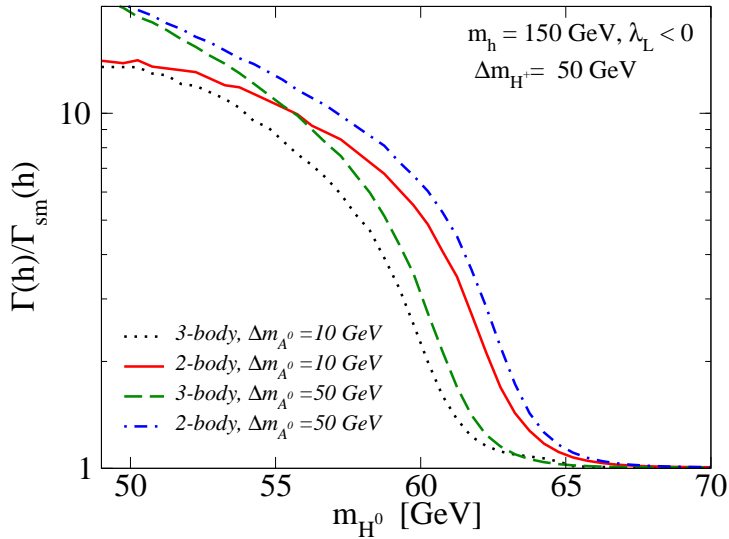}
\caption{Ratio between the higgs branching ratios in the inert doublet model and in the standard model along the viable regions for $m_h=150~\mathrm{GeV}$. $\lambda_L$ was taken to be negative and $\Delta m_{H^\pm}=50$ GeV.\label{fig:hdecays}}
\end{center}
\end{figure}
In the inert doublet model, the higgs boson can decay also into
$H^0H^0$ and $A^0A^0$, increasing the higgs decay width and modifying
its branching ratios. A result that is of great interest for higgs searches at colliders \cite{Cao:2007rm}. Here we simply illustrate how the higgs decay width is
modified when including the three-body final state in the determination of the relic abundance. As before, we assume that  $H^0$ accounts for the dark matter of the Universe.

The  contribution to the higgs decay with from the decay into the inert
scalars is proportional to $\lambda_L^2$, so it will be affected
by the three-body final state $WW^*$ via the
new viable parameter space.
Figure \ref{fig:hdecays} shows, as a
function of $\mH$, the ratio between the higgs decay width in the
inert doublet model and that in the Standard Model, $\Gamma(h)/\Gamma_{sm}(h)$, for
$m_h=150~\gev$.  Obviously, the higgs width  is only
affected for $ m_{H^0} $ below $m_h/2$. 
For the inert doublet model we consider the result
with and without\footnote{Notice that $\Gamma(h)$ and $\Gamma_{sm}(h)$
  includes the $h\rightarrow WW^*$ channel in all the curves} 
taking into account the three-body final state in the
calculation of the relic density, and we set $\dmc=50~\gev$,
$\lambda_L<0$.  First of all, notice that the higgs decay width can
indeed be much larger than what is  predicted by the Standard
Model. Second, the higgs width in the IDM is  slightly smaller once three-body
final states are taken into account --see, for instance, the
difference between the solid and dotted line or between the
dash-dotted and the dashed line. Notice, however, that the region where the deviation from the standard model result is significant, for
$50~\gev<\mH<60~\gev$, is partially excluded by the present direct detection bound from CDMS --see figure \ref{fig:sdd150}. This result emphasizes the importance of doing a consistent analysis of the inert doublet model, one  that simultaneously takes into account all the relevant effects and the different constraints that can be imposed on the model.

\section{Conclusions}
We studied the impact, on the phenomenology of the inert doublet
model, of dark matter annihilation into the three-body final state
$WW^*$. After  analyzing its dependence on  the parameters of the model, the annihilation cross section into $WW^*$,
$\sigma(H^0H^0\to WW^*)$, was shown to dominate the total dark matter
annihilation cross section over a relevant portion of the parameter
space. In consequence, the predicted relic density differs
considerably from that found in earlier works. We examined in detail the dependence of the
inert higgs relic density on the higgs mass and on $\lambda_L$, as well as the possible role of coannihilations. The genuine viable parameter space of the inert doublet model was derived for different values of the higgs mass and of $\dma$, emphasizing the differences with respect to the two-body results.  We also investigated the direct detection cross section of inert higgs dark matter and found that it can be more than two orders of magnitude  smaller than what is  predicted without including the three-body final state. Finally, we briefly consider some implications of these new annihilation processes on the decay width of the higgs boson and on the indirect detection of inert higgs dark matter. Summarizing, the inclusion of the three-body annihilation  is mandatory, as it strongly affects the entire phenomenology of the inert doublet model.
\section*{Acknowledgments} 
L. L. H was partially supported by CICYT through the project FPA2009-09017, by CAM
through the project HEPHACOS, P-ESP-00346, by the PAU (Physics of the accelerating
universe) Consolider Ingenio 2010, by the F.N.R.S. and the I.I.S.N.. C. E. Y. is supported by the \emph{Juan de la Cierva} program of the MICINN of Spain. He acknowledges additional support from the MICINN Consolider-Ingenio 2010 Programme under grant MULTIDARK CSD2009-00064, from the MCIINN under Proyecto Nacional FPA2009-08958, and from the CAM under grant HEPHACOS S2009/ESP-1473.

\appendix
\section{Analytical formula for $\sigma(H^0H^0\to WW^*)$ in the inert doublet model}
\label{ap:1}
Here we provide the total amplitude squared for the process $H^0H^0\to WW^*$, which has  been directly obtained with the Calchep
package~\cite{calchep}. The relevant Feynman diagrams are shown in figure \ref{fig:diainert}. We denote by ${\cal M}_p$ the amplitude for the direct annihilation diagram (left diagram in  figure \ref{fig:diainert}) and by ${\cal M}_s$ the higgs mediated diagram (right diagram). ${\cal M}_{u,t}$ correspond to the $H^+$ mediated diagrams (middle). In the following, $p_1$ and $p_2$ denotes the 4-momentum of the
annihilating $H_0$, $p_3$
and $p_4$ are the momentum of the 2 fermions produced in the decay of
the virtual $W^*$ $(p_{W^*}=p_3+p_4)$ and $p_5$ is the momentum of the real $W$.  
\begin{eqnarray}
 | {\cal M}_p|^2&=&\frac{g^6}{4 M_W^2}\frac{-2 m_{H_0}^2 + M_W^2 + 2 (-p_1.p_2
   + p_1.p_3 + p_1.p_4 + p_2.p_3 + p_2.p_4)}{(((p_3+p_4)^2-M_W^2)^2+(\Gamma_W M_W)^2)}\cr~\cr
|{\cal M}_t|^2&=&\frac{-g^6}{8 M_W^2 D_t}[((m_{H_0}^2 - m_{H^+}^2)^2 - 
   M_W^2 (M_W^2 + 4 (-p_1.p_2 + p_2.p_3 + p_2.p_4)))\cr
 &&\hspace{1.5cm} (2 m_{H_0}^4 + 
   4 p_1.p_3 p_1.p_4 - 
   m_{H_0}^2 (M_W^2 + 2 (-p_1.p_2 + p_1.p_3 + p_1.p_4 + p_2.p_3 +
   p_2.p_4)))]\cr~\cr
D_t&=&{(t-m_{H^+}^2)^2*(((p_3+p_4)^2-M_W^2)^2+(\Gamma_W M_W)^2)}\cr
|{\cal M}_u|^2&=&\frac{-g^6}{8 M_W^2D_u}[-((m_{H_0}^2 - m_{H^+}^2)^2 - 
    M_W^2 (M_W^2 + 4 (-p_1.p_2 + p_1.p_3 + p_1.p_4)))\cr
&&\hspace{1.5cm} (2 m_{H_0}^4 + 
   4 p_2.p_3 p_2.p_4 - 
   m_{H_0}^2 (M_W^2 + 2 (-p_1.p_2 + p_1.p_3 + p_1.p_4 + p_2.p_3 +
   p_2.p_4)))]\cr~\cr
D_u&=&{(u-m_{H^+}^2)^2*(((p_3+p_4)^2-M_W^2)^2+(\Gamma_W M_W)^2)}\cr~\cr
| {\cal M}_s|^2&=&\frac{g^6}{M_W^2D_p}[(m_{H_0}^2 - \mu_2^2)^2 (2 m_{H_0}^4 + M_W^4 + 
   2 (p_1.p_2 - p_1.p_3 - p_2.p_3) (p_1.p_2 - p_1.p_4 - p_2.p_4) \cr
&&\hspace{1.5cm}+ 
   M_W^2 (-3 p_1.p_2 + 2 (p_1.p_3 + p_1.p_4 + p_2.p_3 + p_2.p_4)) \cr
&&\hspace{1.5cm}- 
   m_{H_0}^2 (3 M_W^2 + 2 (-2 p_1.p_2 + p_1.p_3 + p_1.p_4 + p_2.p_3 +
   p_2.p_4)))]\cr~\cr
D_s&=&{((s-m_h^2)^2+ (m_h\Gamma_h)^2)*(((p_3+p_4)^2-M_W^2)^2+(\Gamma_W M_W)^2))}\cr~\cr
2{\cal M}_s{\cal M}_p^\dag&=&\frac{g^6}{M_W^2D_{ps}}[(m_{H_0} - \mu_2)
  (m_{H_0} + \mu_2) \cr
&&\hspace{1.5cm}(2 m_{H_0}^2 - M_W^2 - 
   2 (-p_1.p_2 + p_1.p_3 + p_1.p_4 + p_2.p_3 +
   p_2.p_4))]\cr~\cr
D_{sp}&=&{(s-m_h^2)*(((p_3+p_4)^2-M_W^2)^2+(\Gamma_W M_W)^2)}\cr
2{\cal M}_s{\cal M}_t^\dag&=&\frac{g^6}{2M_W^2D_{st}}[(m_{H_0}^2 - \mu_2^2) (2 m_{H_0}^6  \cr
  &&\hspace{1.5cm}+ m_{H_0}^4 (-2 m_{H^+}^2 + M_W^2 + 4 p_1.p_2 - 
      2 (p_1.p_3 + p_1.p_4 + p_2.p_3 + p_2.p_4))  \cr
  &&\hspace{1.5cm} + m_{H^+}^2 (M_W^2 p_1.p_2 - 
      2 (p_1.p_2^2 + p_1.p_4 p_2.p_3 \cr
&&\hspace{1.5cm}+ p_1.p_3 (2 p_1.p_4 + p_2.p_4) - 
         p_1.p_2 (p_1.p_3 + p_1.p_4 + p_2.p_3 + p_2.p_4))) \cr
&&\hspace{1.5cm}+ 
   M_W^2 (M_W^2 p_1.p_2 + 
      2 (-p_1.p_2^2 + 2 p_1.p_3 p_1.p_4 \cr
&&\hspace{1.5cm}- p_1.p_4 p_2.p_3 - p_1.p_3 p_2.p_4 +
          p_1.p_2 (p_1.p_3 + p_1.p_4 + p_2.p_3 + p_2.p_4))) \cr
&&\hspace{1.5cm}+ 
   m_{H_0}^2 (-M_W^4 - 
      M_W^2 (p_1.p_2 + 2 (p_1.p_3 + p_1.p_4 + p_2.p_3 + p_2.p_4)) \cr
&&\hspace{1.5cm}+ 
      2 (p_1.p_2^2 + p_1.p_4 p_2.p_3 + p_1.p_3 (2 p_1.p_4 + p_2.p_4)\cr
&& \hspace{1.5cm}- 
         p_1.p_2 (p_1.p_3 + p_1.p_4 + p_2.p_3 + p_2.p_4)) \cr
&&\hspace{1.5cm}+ 
      m_{H^+}^2 (M_W^2 + 
         2 (-2 p_1.p_2 + p_1.p_3 + p_1.p_4 + p_2.p_3 + p_2.p_4)))) ]\cr~\cr
D_{st}&=&{(s-m_h^2)*(((p_3+p_4)^2-M_W^2)^2+(\Gamma_W M_W)^2)*(t-m_{H^+}^2)}\cr~\cr
2{\cal M}_s{\cal M}_u^\dag&=&\frac{-g^6}{2M_W^2D_{su}}[-(m_{H_0}^2 -
  \mu_2^2) (2 m_{H_0}^6 \cr
&&\hspace{1.5cm}+ 
   m_{H_0}^4 (-2 m_{H^+}^2 + M_W^2 + 4 p_1.p_2 - 
      2 (p_1.p_3 + p_1.p_4 + p_2.p_3 + p_2.p_4)) \cr
&&\hspace{1.5cm}+ 
   m_{H^+}^2 (M_W^2 p_1.p_2 - 
      2 (p_1.p_2^2 + p_1.p_4 p_2.p_3\cr
&&\hspace{1.5cm} + (p_1.p_3 + 2 p_2.p_3) p_2.p_4 - 
         p_1.p_2 (p_1.p_3 + p_1.p_4 + p_2.p_3 + p_2.p_4))) \cr
&&\hspace{1.5cm}+ 
   M_W^2 (M_W^2 p_1.p_2 + 
      2 (-p_1.p_2^2 - p_1.p_4 p_2.p_3 \cr
&&\hspace{1.5cm}- p_1.p_3 p_2.p_4 + 2 p_2.p_3 p_2.p_4 +
          p_1.p_2 (p_1.p_3 + p_1.p_4 + p_2.p_3 + p_2.p_4))) \cr
&&\hspace{1.5cm}+   m_{H_0}^2 (-M_W^4 - 
      M_W^2 (p_1.p_2 + 2 (p_1.p_3 + p_1.p_4 + p_2.p_3 + p_2.p_4)) \cr
&&\hspace{1.5cm}+ 
      2 (p_1.p_2^2 + p_1.p_4 p_2.p_3 + (p_1.p_3 + 2 p_2.p_3) p_2.p_4
      \cr
&&\hspace{1.5cm}- 
         p_1.p_2 (p_1.p_3 + p_1.p_4 + p_2.p_3 + p_2.p_4)) \cr
&&\hspace{1.5cm}+ 
      m_{H^+}^2 (M_W^2 + 
      2 (-2 p_1.p_2 + p_1.p_3 + p_1.p_4 + p_2.p_3 + p_2.p_4))))]\cr~\cr
D_{su}&=&{(s-m_h^2)*(((p_3+p_4)^2-M_W^2)^2+(\Gamma_W M_W)^2)*(u-m_{H^+}^2)}\cr~\cr
2{\cal M}_p{\cal M}_t^\dag&=&\frac{g^6}{4 M_W^2D_{pt}}[-2 m_{H_0}^4 -
  M_W^2 p_1.p_2  
+ m_{H_0}^2 (M_W^2 + 2 (p_1.p_3 + p_1.p_4 + p_2.p_3 + p_2.p_4)) \cr
&&\hspace{1.5cm}+ 
 2 (p_1.p_2^2 + p_1.p_4 p_2.p_3 + p_1.p_3 (-2 p_1.p_4 + p_2.p_4) \cr
&&\hspace{1.5cm}- 
    p_1.p_2 (p_1.p_3 + p_1.p_4 + p_2.p_3 + p_2.p_4))]\cr~\cr
D_{pt}&=&{(((p_3+p_4)^2-M_W^2)^2+(\Gamma_W M_W)^2)*(t-m_{H^+}^2)}\cr~\cr
2{\cal M}_p{\cal M}_u^\dag&=&\frac{-g^6}{4 M_W^2D_{pu}}[2 m_{H_0}^4 + M_W^2 p_1.p_2 - 
 m_{H_0}^2 (M_W^2 + 2 (p_1.p_3 + p_1.p_4 + p_2.p_3 + p_2.p_4))\cr
&& \hspace{1.5cm}+ 
 2 (-p_1.p_2^2 - p_1.p_4 p_2.p_3 - p_1.p_3 p_2.p_4 + 2 p_2.p_3 p_2.p_4 \cr
&&\hspace{1.5cm}+ 
    p_1.p_2 (p_1.p_3 + p_1.p_4 + p_2.p_3 +
    p_2.p_4))]\cr~\cr
D_{pu}&=&{(((p_3+p_4)^2-M_W^2)^2+(\Gamma_W M_W)^2)*(u-m_{H^+}^2)}\cr~\cr
2{\cal M}_t{\cal M}_u^\dag&=&\frac{-g^6}{4 M_W^2 D_{tu}}[-(m_{H_0}^4 +
  m_{H^+}^4 - 2 m_{H_0}^2 (m_{H^+}^2 - 2 M_W^2) \cr
&&\hspace{1.5cm}- 
    M_W^2 (M_W^2 + 
       2 (p_1.p_3 + p_1.p_4 + p_2.p_3 + p_2.p_4))) (2 m_{H_0}^2 p_1.p_2 - 
   M_W^2 p_1.p_2 \cr
&&\hspace{1.5cm}+ 
   2 (p_1.p_2^2 + p_1.p_4 p_2.p_3 + p_1.p_3 p_2.p_4 - 
      p_1.p_2 (p_1.p_3 + p_1.p_4 + p_2.p_3 + p_2.p_4)))]\cr~\cr
D_{tu}&=&{(t-m_{H^+}^2)*(((p_3+p_4)^2-M_W^2)^2+(\Gamma_W M_W)^2)*(u-m_{H^+}^2)}\nonumber
\end{eqnarray}
The cross section $\sigma(H^0H^0\to WW^*)$ can then be obtained, from the total amplitude,  in the usual way.

\section{Interference effects around the higgs resonance}
\label{ap:2}
Around the higgs resonance, cancellations between different contributions clearly appear in  $\sigma v(\text{3-body})$.  They take place at $m_{H_0}<m_h/2$ for $ \lambda_L>0$ and at $m_{H_0}>m_h/2$ for 
$ \lambda_L<0$. This fact can be easily understood by looking at the square of
the sum of the two relevant amplitudes, which are the one of the point
like diagram ${\cal M}_p$ and the one of the higgs mediated diagrams ${\cal M}_s$. They can be written as
\begin{eqnarray}
   i{\cal M}_p&=&-\frac{g^2}{2} {\cal O}_{W\mu} {\cal
     O}_{W^*}^\mu\label{eq:Mp}\cr
 i{\cal M}_s&=&-\frac{\lambda_L v^2 g^2}{s-m_h^2-i m_h\Gamma_h} {\cal O}_{W\mu} {\cal O}_{W^*}^\mu\label{eq:Ms}
\end{eqnarray}
where, in a similar way to \cite{Chen:1998dp}, we have defined ${\cal O}_{W\mu}$ as the
wave function of the outgoing real $W$ and ${\cal O}_{W^* \mu}$ as the
contribution from the virtual gauge boson and the two outgoing fermions:
\begin{eqnarray}
 {\cal O}_{W^* \mu}&=&\frac{-i
   g_{\mu\nu}}{(p_3+p_4)^2-M_W^2-i\Gamma_W M_W}\frac{g}{\sqrt{2}}\bar
 u_3 \gamma^\nu v_4\label{eq:OWs}\,.
\end{eqnarray}
For the square of their sum we obtain
\begin{equation}
  |i{\cal M}_p+ i{\cal
  M}_s|^2\simeq \frac{g^4}{4}\frac{|{\cal O}_{W\mu} {\cal
      O}_{W^*}^\mu|^2}{{(s-m_h^2)^2+
      m_h^2\Gamma_h^2}} [(s-m_h^2)+2\lambda_L v^2]^2
\end{equation}
Since  $s\simeq 4 m_{H_0}^2$ at low velocities,   a negative interference between the two terms  should occur for $m_{H_0}<m_h/2$ if $ \lambda_L>0$ and for $m_{H_0}>m_h/2$ if $ \lambda_L<0$. That explains why the 3-body annihilation cross section is larger for $\lambda_L<0$ than for $\lambda_L>0$ if $m_{H^0}<m_h/2$ but it is the other way around for $m_{H^0}>m_h/2$. Moreover, a  cancellation between these two contributions takes place  at
$|m_{H_0}-m_h/2|\simeq 1/2 |\lambda_L| v^2/m_h$. For $m_h=120,150$
GeV and $\lambda_L=10^{-2}$, that position correspond to $|m_{H_0}-m_h/2|\simeq 2.5, 2.0$ GeV, in good agreement
with figures \ref{fig:sigmav},~\ref{fig:svlambda} and~\ref{fig:svsignlam}, which were obtained integrating
numerically the cross section including all the contributions displayed
in appendix~\ref{ap:1}.   

\bibliography{scalar}{}
\bibliographystyle{hunsrt}

\end{document}